\documentclass[a4paper,10pt,reqno, number]{amsart}

\usepackage[utf8]{inputenc}
\usepackage[    hyperref, backend=biber, doi=false,url=false,sorting=none,backref=true]{biblatex}

\usepackage{amsmath}
\usepackage{amssymb}
\usepackage{color}

\usepackage{graphicx}
\usepackage{epstopdf}

\newcommand{\coqui}[1]{{#1}}

\usepackage[english]{babel}
\bibliography{bibliografia}

\newcommand{\repre}{S}
\newcommand{\argo}{\left(\beta\momenta\right)}
\newcommand{\pressure}{\mathcal{F}}
\newcommand{\energy}{\mathcal{E}}
\newcommand{\momenta}{k_n}
\newcommand{\momentatwo}[1]{k_{#1}}

\title{Casimir effect in Snyder Space}

\author{S.~A.~Franchino-Vi\~nas}
\address{Departamento de F\'isica, Facultad de Ciencias Exactas
Universidad Nacional de La Plata, C.C.\ 67 (1900), La Plata, Argentina.}

\address{Institut für Theoretische Physik, Universität Heidelberg, D-69120 Heidelberg, Germany.}

\address{Theoretisch-Physikalisches Institut, Friedrich Schiller Universit\"at Jena, Max Wien Platz 1, 07743 Jena, Germany.}
\email{sa.franchino@uni-jena.de}

\author{S. Mignemi}
\address{Dipartimento di Matematica e Informatica, Università di Cagliari, viale Merello 92, \\09123 Cagliari, Italy, and \newline
INFN, Sezione di Cagliari, Cittadella Universitaria, 09042 Monserrato, Italy.}
\email{smignemi@unica.it}

\begin{document}
\begin{abstract}
We show that two indistinguishable aspects of the divergences occurring in the Casimir effect, namely
the divergence of the energy of the higher modes and the non-com\-pact\-ness of the momentum space,
get disentangled in a given noncommutative setup.
To this end, we consider a scalar field between two parallel plates in an anti-Snyder space.
\coqui{Additionally, the large mass decay in this noncommutative setup
is not necessarily  exponential.}

\end{abstract}

\maketitle

\section{Introduction}
Many  questions  in the realm of particle physics have been answered by the Quantum Theory of Fields (QFT). This theory has offered us many results, some of them of fundamental simplicity and beauty. Among these one can probably include the Casimir effect. First predicted by Casimir in 1948 \cite{Casimir:1948dh} and experimentally confirmed a decade after by Sparnaay \cite{Sparnaay:1957}, it predicts that, being an (infinite) sum of harmonic oscillators, fields in QFT have a vacuum energy that depends on the geometry of the space. Like many other quantities in QFT this sum is divergent and should be regularized in order to obtain physical results. It also encodes a deep connection between geometry and QFT, since the energy of every single oscillator depends on it. Since the literature is really vast, we refer the reader to some of the numerous reports and books on the subject \cite{ Plunien:1986ca, Milton:2001yy, Bordag:2009zzd}. Here we shall only mention some of the latest results.

The fields of application of the Casimir effect are innumerable. Among the most interesting possibilities are its applications in condensed matter physics, where one can mimic the behaviour of several materials through the inclusion of boundary conditions, considering special topologies or effects such as inhomogeneities \cite{Asorey:2013wca, Munoz-Castaneda:2013yga, Vinas:2010ix, Bellucci:2019ybj,Fosco:2019lmw}. Also some intriguing facts related to the observation of negative entropy for finite temperature vacuum energies have been discussed in the last years \cite{Bordag:2019vrw}.

More important to us  are its applications to theories Beyond the Standard Model (BSM), which run from the consideration of the effect in curved spaces \cite{BezerradeMello:2017nyo, Saharian:2020uiu} to the possible implications on neutrino oscillations \cite{Blasone:2018obn} or more general scenarios like brane-worlds, extra dimensions, scale-invariant models and  generalized uncertainty principle \cite{Bellucci:2019viw, Mattioli:2019xgl, Frassino:2011aa}. From the experimental side, many constraints related to possible modifications of Newtonian gravity  have been obtained \cite{Kuzmin:1982ei} in the last decades; the latest results fix for example stringent bounds to the axion mass and coupling \cite{Klimchitskaya:2020cnr}.

Here we will focus on still another BSM scenario, viz.~noncommutative QFT, one of the most prominent and studied candidates for an effective field theory of Quantum Gravity. The main idea behind this theory is that the quantum nature of geometry could first manifest through the existence of noncanonical commutation relations between position and momentum operators, which could contribute to regularize the usual divergences in QFT.

Of course the effect of vacuum energy  has been widely studied in the context of noncommutative field theories of scalar fields for various models, using different methods: for the Moyal torus and cylinder \cite{Chaichian:2001pw}, for the Moyal (hyper)plane from a heuristic point of view \cite{Casadio:2007ec, Fosco:2007tn}, in the case of Snyder spaces with the use of heat-kernel techniques \cite{Mignemi:2017yhd} and for $\kappa$-Minkowski space adopting the energy-momentum tensor approach \cite{Harikumar:2019hzq}.

In this paper, we consider the Casimir energy density  for a  scalar field  theory confined to a slab between two parallel plates in anti-Snyder space. This choice is  motivated by the fact that in Snyder space the Lorentz symmetry is undeformed, in contrast with other noncommutative setups.  Moreover, a formulation of a QFT on both its flat \cite{Meljanac:2017grw, Franchino-Vinas:2018jcs} and curved space \cite{Franchino-Vinas:2019lyi,Franchino-Vinas:2019nqy} versions have been recently pursued, evidentiating several interesting results.  However, to our knowledge this is the first study of QFT in a bounded region of Snyder space.

We will provide a short review of the Snyder geometry in Sec.~\ref{sec:snyder}. Then, in Sec.~\ref{sec:spectrum},
we will derive the spectrum for the geometry determined by two parallel plates in anti-Snyder space, by means of a suitable confining potential.
This result will be used in Sec. \ref{sec:casimir} in order to derive an expression for the Casimir energy of a slab in $\mathcal{M}=\mathbb{R}\times (\text{anti-Snyder})_D$.
{We will show that there are two possible interpretations, depending on the nature of the involved cutoff $\Lambda$. In the case where there exists a natural UV-cutoff $\Lambda<\beta^{-1}$, we will consider the derivation of a regularized pressure in $D=1$ and $D=3$ dimensions in Sec. \ref{sec:casimir_d=1} and Sec. \ref{sec:casimir_d=3} respectively. In absence of such a natural UV-cutoff, $\Lambda$ can be interpreted as a cutoff for distant modes in momentum space; this geometric point of view will be examined in Sec. \ref{sec:noncommutative_casimir}}. Finally, we will discuss our results in Sec. \ref{sec:conclusions}.

\section{The Snyder model}\label{sec:snyder}

Here we shall summarize the main properties of the Euclidean $D-$dimensional anti-Snyder model that will be used in the following.
The model is based on the following commutation relations between the operators of position ($\hat x_i$), momentum ($\hat p_i$),
and Lorentz generators ($\hat J_{ij}=\hat x_i\hat p_j-\hat x_j\hat p_i$) \cite{Mignemi:2011gr}:
\begin{eqnarray}\label{Snydercomm}
&&[\hat J_{ij},\hat J_{kl}]=i\left(\delta_{ik}\hat J_{jl}-\delta_{il}\hat J_{jk}-\delta_{jk}\hat J_{il}+\delta_{lj}\hat J_{ik}\right),\cr
&&[\hat J_{ij},\hat p_k]=i\left(\delta_{ik}\hat p_j-\delta_{jk}\hat p_i\right),\qquad[\hat J_{ij},\hat x_k]=i\left(\delta_{ik}\hat x_j-\delta_{jk}\hat x_i\right),\cr
&&[\hat x_i,\hat p_j]=i\left(\delta_{ij}-\beta^2\hat p_i\hat p_j\right),\qquad[\hat x_i,\hat x_j]=-i\beta^2\hat J_{ij},\qquad[\hat p_i,\hat p_j]=0,
\end{eqnarray}
where $\beta$ is a constant of order $1/M_P$, with $M_P$ the Planck mass, and $i,j=1,\dots,D$.
These commutation relations include those of the Lorentz algebra, with its standard action on phase space,
and a deformation of the Heisenberg algebra.
In this paper we shall consider the anti-Snyder model, but a variant called Snyder model, with opposite sign of $\beta^2$, is often considered. They differ in several respects. In particular, the spectrum of the square of momentum is continuous but bounded in anti-Snyder space, $\hat p^2<1/\beta^2$, while the opposite holds in the other case.
Geometrically the anti-Snyder  momentum space is an hyperbolic space.

Several representations of the commutation relations \eqref{Snydercomm} on a Hilbert space are possible: the original one, which will be referred to as Snyder representation \cite{Snyder:1946qz,Lu:2011fh}, is defined by the operators
\begin{align}\label{eq:realization2}
 \hat{p}_i=p_i,\quad \hat{x}_i=i\left(\delta_{ij}-\beta^2p_ip_j\right)  \frac{\partial}{\partial p_j}.
\end{align}
acting on a Hilbert space of functions $\psi(p)$ with measure  $d\mu=\frac{d^Dp}{(1-\beta^2p^2)^{(D+1)/2}}$ \cite{Lu:2011fh}.

A different  realization was introduced in \cite{Mignemi:2011gr}:
\begin{align}\label{eq:realization1}
 \hat{p}_i=\frac{p_i}{\sqrt{1+\beta^2{p^2}}},\qquad \hat{x}_i= i \sqrt{1+\beta^2p^2} \frac{\partial}{\partial p_i}.
\end{align}
The measure on the Hilbert space is in this case  $d\mu=\frac{d^Dp}{\sqrt{1+\beta^2p^2}}$.
In the following we shall use the latter realization, because it leads to simpler calculations. The two are of course related by a unitary transformation.

\section{The spectrum of a confined particle in anti-Snyder space}\label{sec:spectrum}
In order to compute the Casimir energy of a scalar field in anti-Snyder space we will follow an approach similar to the one of the original derivation by Casimir, i.e. we will consider the spectrum of the one-loop quantum fluctuations of the confined field and sum over all the possible modes.

However, in a noncommutative space the derivation of the spectrum is not straightforward, since the imposition of boundary conditions is hindered by the granularity of the background spacetime.
We will avoid this issue by introducing walls of finite potential $V$ situated on the hyperplanes\footnote{We will denote the direction perpendicular to the plates with the subscript $\perp$, while for the remaining parallel $D-1$ dimensions we will use the symbol $\parallel$.} $x_{\perp}=\pm L$ and then taking the limit $V\to\infty$, namely, we will consider the eigenstates of the operator
\begin{align}\label{eq:hamiltonian}
H_V= \hat{p}^2+ V H(\hat{x}_{\perp}-L)+ V H(-\hat{x}_{\perp}-L)
\end{align}
for infinite $V$, with $H(\cdot)$ the Heaviside function.

We shall work in the representation \eqref{eq:realization1}, in which the eigenstates of momenta operators take the form\footnote{We shall often suppress the vector index in the notation.}
\begin{align}
 \phi_{q}(p)=\sqrt{1+\beta^2q^2}\delta (p-q),
\end{align}
with eigenvalues $\frac{q_i}{\sqrt{1+\beta^2q^2}}$, which are normalized such that the completeness relation gives rise to the usual covariant delta function in curved space,
\begin{align}
\int \frac{d^D q}{\sqrt{1+\beta^2q^2}} \phi_q(p) \phi_q^*(p')&=\sqrt{1+\beta^2p^2}\delta^D(p-p')
\end{align}

Remarkably, the existence of eigenfunctions $ \psi_{x_i}(p)$  of the components of the position operators has passed unnoticed in the literature,
\begin{align}
 \psi_{x_i}(p)=e^{-i \frac{x_i}{\beta} \text{arcth}\left(\frac{\beta p_i}{\sqrt{1+\beta^2p^2}}\right)},\quad  \hat{x}_i\psi_{x_i}(p)=x_i \psi_{x_i}(p).
 \end{align}
 Of course, since the components of the position operator do not commute among themselves, these states cannot form a basis of the Hilbert space. However,
a complete basis of generalized states can be obtained, parametrized by the quantum numbers $x_{\perp}$ and $q_{\parallel}$, i.e. with a given position $x_{\perp}$ in a  fixed direction, and momentum components $q_{\parallel}$ in the orthogonal directions:
\begin{align}
 \psi_{x_{\perp},q_{\parallel}}(p):= \frac{1}{\sqrt{2\pi}}\psi_{x_{\perp}}(p) \delta(p_{\parallel}-q_{\parallel}).
\end{align}
These states are orthonormal, in the sense that their scalar product in momentum space is given by
 \begin{align}
  \left( \psi_{x_{\perp},q_{\parallel}} ,\psi_{y_{\perp},k_{\parallel}}\right)
  &=\delta(k_{\parallel}-q_{\parallel}) \delta(x_{\perp}-y_{\perp}).
 \end{align}

 With these ingredients, we are ready to compute the spectrum of $H_{\infty}$. Indeed, the eigenfunctions of the momenta can be thought as eigenfunctions of $H_V$ in the regions of constant $V$. The key idea is that, in the different regions, one can  combine the left- and right-travelling eigenfunctions in the direction $x_{\perp}$ and glue  them together, since they have the same energy. One can think of this as considering the projectors of the proposed solution into position eigenstates and asking for continuity. For example, if we call the solution in the whole space $\Psi_{q}(p)$, in the middle region we would obtain
\begin{align}
 \Psi_q(p')=\int_{-L}^{L} d{x_{\perp}} \int \frac{dp^4}{\sqrt{1+\beta^2p^2}} \psi^*_{x_{\perp},q_{\parallel}}(p) \left(A_{q}\phi_q(p)+B_{q} \phi_{-q_{\perp},q_{\parallel}}(p)\right)\psi_{x_{\perp},q_{\parallel}}(p').
\end{align}
In the limit of infinite potential, continuity requires that the projection into eigenstates of $\hat x_{\perp}$ with eigenvalue $\pm L$ should vanish, viz.
\begin{align}
\left(\psi_{\pm L,q_{\parallel}} ,\; A_{q}\phi_q +B_{q} \phi_{-q_{\perp},q_{\parallel}} \right)=0.
\end{align}
Therefore, we obtain as usual a system of two equations whose compatibility entails the quantization of the energies. This condition can be written as
\begin{align}
  \sin\left( \frac{2L}{\beta} \text{arcth}\left( \frac{\beta q_{\perp}}{\sqrt{1+\beta^2q^2}}\right) \right)=0,
\end{align}
from which one can obtain the spectrum of the momenta $q_{\perp}$,
\begin{align}\label{eq:spectrum}
 \beta^2 q_{\perp,n}^2 &=\sinh^2\left(\momenta \beta\right) \left(1+\beta^2q_{\parallel}^2\right),\quad n\in\mathbb{N}^{+}.
\end{align}
Notice that this result is well-behaved in the commutative limit of vanishing $\beta$, from which one can recover the known commutative quantization rule
\begin{align}
  q_{\perp,n} \xrightarrow{\beta\rightarrow 0 } \frac{n\pi}{2L}=:\momenta.
\end{align}


\section{The Casimir energy}\label{sec:casimir}
One can consider a scalar quantum field theory built on a $D+1$ dimensional manifold given by $\mathcal{M}_{D+1}=\mathbb{R}\times (\text{anti-Snyder})_D$, or even its restriction to the slab described by the imposition of the previous Dirichlet boundary conditions. The fact that we have chosen the time to be commutative avoids the well-known unitarity problems that arise in some noncommutative theories.  In this manifold, the wave equation for a field of mass $m$ will be
\begin{align}
 (\partial_t^2+\hat p^2+m^2)\,\phi =0,
\end{align}
where $\hat p^2$ is the generalized Laplacian of $D$-dimensional Snyder space.
If we consider states of definite energy $\omega$, denoted by $\phi=e^{i\omega t}\, \phi_{\omega}$  which are eigenstates of the operator $\hat p^2$ with eigenvalue $p^2$,
\begin{align}\label{eq:dispersion}
 \omega^2=p^2+m^2,
\end{align}
the dispersion relation can be readily obtained replacing \eqref{eq:spectrum} in \eqref{eq:dispersion}: written in terms of the auxiliary variables $q$, it takes the form of a deformed dispersion relation,
\coqui{\begin{align}\label{eq:dispersion2}
 \begin{split}
\omega_{q_{\parallel},n}^2&= \frac{q_{\parallel}^2+q_{\perp,n}^2}{1+\beta^2 \left(q_{\parallel}^2+q_{\perp,n}^2\right)} +m^2\\
 &= \frac{q_{\parallel}^2+\beta^{-2}\sinh^2\left(\beta\momenta\right) \left(1+\beta^2q_{\parallel}^2\right) }{1+\beta^2q_{\parallel}^2+\sinh^2\left(\beta\momenta \right) \left(1+\beta^2q_{\parallel}^2\right)} +m^2.
 \end{split}
\end{align}
}
We can then obtain the formula for the Casimir energy  by summing over all the available modes, i.e.~taking a sum over
the discrete index $n$ corresponding to the direction perpendicular to the plates and integrating the continuous variables representing the parallel directions.
Dropping the $\parallel$ symbol to simplify the expression and calling $\Omega_{D}$ the hypersurface of the unit $D$-sphere,
we obtain an expression for the energy density $\energy$ per unit area in the parallel directions to the plate ($x_{\parallel}$)
\begin{align}\label{eq:casimir}
 \energy
 &= \frac{\Omega_{D-2}}{2} \sum_{n=1}^{\infty~}\int_0^{\infty} \frac{dq}{(2\pi)^{D-1}}\,q^{D-2} \sqrt{\frac{q^2}{1+\beta^2q^2}+\frac{\tanh^2\left(\beta\momenta\right)}{\beta^2(1+\beta^2q^2)}+m^2}.
\end{align}

It is important to notice that in this expression the contribution of the measure introduced short after eq.~\eqref{eq:realization1}, does not appear explicitly. In fact, it is cancelled by other contributions coming from the normalization of the modes. This will be of crucial importance in the discussion of the realization independence of the Casimir energy density.

The correctness of equation \eqref{eq:casimir} can be checked by showing that in the $L\rightarrow \infty$ limit one obtains the correct result: \coqui{indeed, considering $\tilde n:= n/L$, which will become continuous in the large-$L$ limit, and then changing variables to
\begin{align}\label{eq:change_variables}
z=\beta^{-1}\sinh\left(\beta L\momentatwo{\tilde n} \right) \sqrt{1+\beta^2q^2}
\end{align}
we obtain
\begin{align}\begin{split}\label{eq:energy_coordinates1}
 \energy  &=\frac{\Omega_{D-2}}{2} L \int_0^{\infty} d{\tilde n} \int_0^{\infty} \frac{dq\, q^{D-2}}{(2\pi)^{D-1}}\, \sqrt{\frac{q^2}{1+\beta^2q^2}+\frac{\tanh^2\left(\beta L\momentatwo{{\tilde n}} \right)}{\beta^2(1+\beta^2q^2)}+m^2}+\mathcal{O}(L^0)\\
 &=\frac{L \Omega_{D-2}}{\pi (2\pi)^{D-1}}  \int_0^{\infty}\int_0^{\infty} \frac{dz dq\,q^{D-2}}{\sqrt{1+\beta q^2+\beta^2z^2}}\, \sqrt{\frac{q^2+z^2}{1+\beta^2q^2+\beta^2 z^2}+m^2}+\mathcal{O}(L^0),
\end{split}
\end{align}}in agreement with the expression given in \cite{Mignemi:2017yhd} for the vacuum energy density in \coqui{the Snyder space in the absence of plates}.

Let us go back to the energy density \eqref{eq:casimir}. In the present form it is divergent. Even if  this is not surprising, since the same behavior occurs in the commutative case, one could have expected the noncommutativity to regularize the divergences. For example, for models confined to a compact manifold, the generalized uncertainty principle for noncommuting coordinates limits the number of modes to be finite; for instance, this is  the case of the fuzzy disc and of the fuzzy sphere \cite{Madore:1991bw,Lizzi:2003ru, Falomir:2013vaa, Franchino-Vinas:2018gbv}. However, in the present case the field is defined on a noncompact space, and therefore the number of states is not constrained (in fact, it is infinite).

However, as discussed above, the momenta are bounded by $p^2<\beta^{-2}$. On the one side, this means that the expected divergence should be somewhat milder than in the commutative case. On the other side, this fact prevents us from using some mathematical regularizations like the $\zeta$-regularization \cite{Elizalde:2007du} or dimensional regularization \cite{Bollini:1972ui, tHooft:1972tcz}. Nevertheless, if we first perform a transformation of variables $p=\frac{q}{\sqrt{1+\beta^2q^2}}$, that brings us back to the physical value $p$ of the momentum, we get
\coqui{\begin{align}\label{eq:energy_coordinates2}
 \begin{split}
\energy
 &=\frac{\Omega_{D-2}}{2(2\pi)^{D-1}} \sum_{n=1}^{\infty}\int_0^{1/\beta} \frac{dp\,p^{D-2}}{(1-\beta^2p^2)^{D/2+1/2}}\, \\
 &\hspace{4cm} \times \sqrt{\frac{p^2-\beta^{-2}}{\cosh^2\left(\beta\momenta\right) }+\beta^{-2} +m^2}.
\end{split}
\end{align}
}
Writing the energy density  in this form, a dimensional regularization seems to be possible even if it is not clear how one could tackle the divergence in the discrete sector. Therefore, we will introduce a physical cutoff on the momentum space. The interested reader could see \cite{Visser:2016ddm} for an interesting discussion on regularization vs. renormalization.

In the following sections we will focus on the $D=1$ and $D=3$ cases, since the former is the easiest one, while the other is the most relevant one for our physical world.

\section{The Casimir force in D=1}\label{sec:casimir_d=1}
Let us first consider  as a toy mode the case of two-dimensional spacetime. The formal expression for the \coqui{ vacuum energy density} is given by
\begin{align}\label{eq:energy_D=1}
 \energy_{D=1}&=\frac{1}{2}\sum_{n=1}^{\infty} \sqrt{\frac{\tanh^2\left(\beta\momenta\right)}{\beta^2}+m^2}.
\end{align}
As customary, one can consider the \coqui{vacuum pressure} $\pressure$, i.e.~the force per unit parallel area applied to the plates, by taking the derivative of the energy density with respect to the distance between them,
\begin{align}\label{eq:force_D=1}
 \pressure_{D=1}=-\frac{1}{2}\partial_L \energy_{D=1}= \sum_{n=1}^{\infty} \frac{\momenta}{4 L}\frac{\tanh\left(\beta\momenta\right)}{\cosh^2\argo\sqrt{\tanh^2\left(\beta\momenta\right)+\coqui{\beta^2 m^2}}}.
\end{align}
This expression  is convergent as it stands. \coqui{However, it should be noticed that it tends to a nonvanishing constant for large distances $L$. Indeed, this  corresponds to the pressure felt by one plate in a single-plate configuration. A careful analysis shows thus that the correct expression for the Casimir pressure, which should involve only the interaction among the plates, is given by eq.  \eqref{eq:force_D=1} after the subtraction of its large $L$ limit-- after a rescaling in the integral we get
\begin{align}\label{eq:force^C_D=1}
 \begin{split}
\pressure^{(C)}_{D=1}:&=\sum_{n=1}^{\infty} \frac{\momenta}{4 L}\frac{\tanh\left(\beta\momenta\right)}{\cosh^2\argo\sqrt{\tanh^2\left(\beta\momenta\right)+\coqui{\beta^2 m^2}}}\\
 &\hspace{2cm}-\frac{1}{2\pi \beta^2}\int_{0}^{\infty} dx  \frac{x \tanh\left( x\right)}{\cosh^2 x \sqrt{\tanh^2\left(x \right)+\coqui{\beta^2 m^2}}}.  
 \end{split}
\end{align}
}\noindent
\noindent \coqui{ Even if in the massive case we are not able to find a closed expression for  eq. \eqref{eq:force^C_D=1}, it is easy to show that the result is attractive, i.e. $\pressure^{(C)}_{D=1}<0$ for every possible choice of the involved parameters. Moreover, in order to extract further information one can evaluate the Casimir pressure numerically.}

\begin{figure}
\begin{center}
 \hspace{-1cm}\begin{minipage}{0.49\textwidth}
 \includegraphics[width=1.1\textwidth]{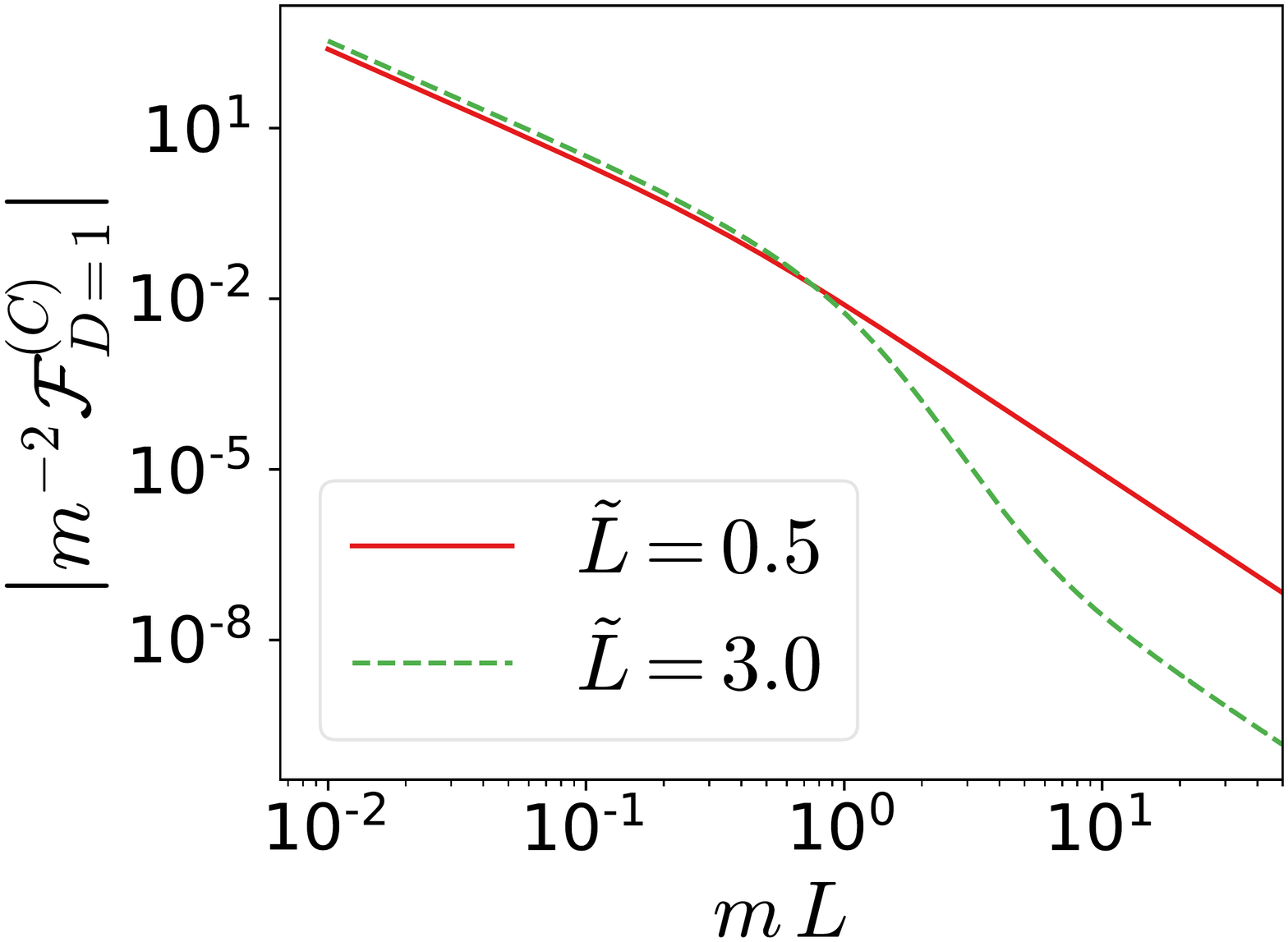}
 \end{minipage}
 \hspace{0.5cm}\begin{minipage}{0.49\textwidth}
 \includegraphics[width=1.1\textwidth]{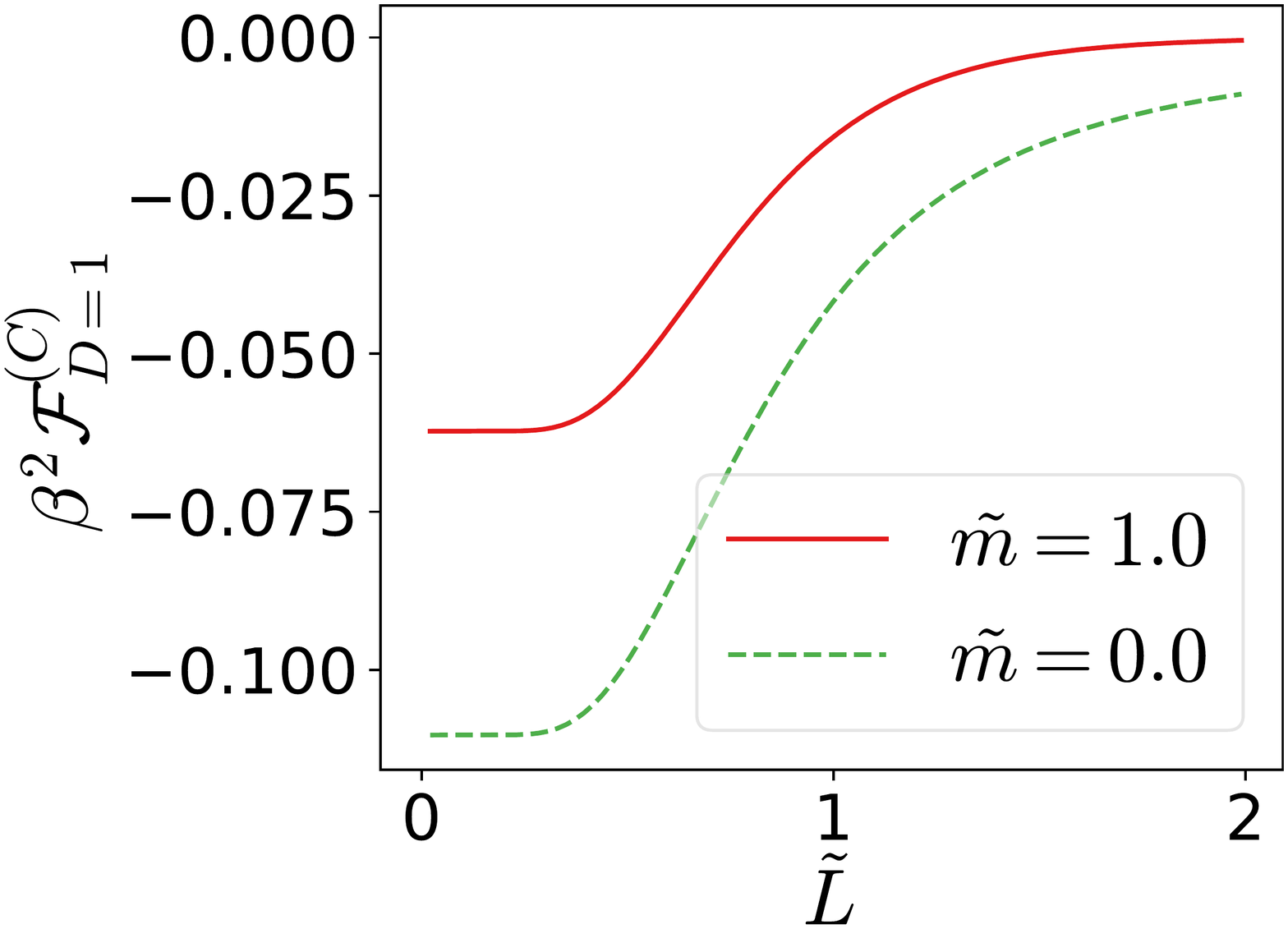}
 \end{minipage}
 \caption{\coqui{The log-log plot on the left panel corresponds to \coqui{$\left\vert m^{-2}\pressure_{D=1}\right\vert$} as a function of $mL$, for $\tilde L=0.5$ (red continuous line) and $\tilde L = 3$ (green dashed line). On the right panel, the plot of $\beta^2\pressure_{D=1}$ as a function of $\tilde L$ is shown, for $\tilde m=1$  (red continous line) and   $\tilde m=0$ (green dashed line).}}
 \label{fig.forced1}
 \end{center}
\end{figure}

\coqui{Notice first of all that, with respect to the commutative case, the present one is richer since we have three dimensionful} parameters. \coqui{In order to simplify the notation, let us introduce the dimensionless parameters $\tilde m= \beta m$ and $\tilde L= \beta^{-1}L$. One can then consider for example the behaviour of the pressure in units of mass, as a function of $mL$ for a given $\tilde L$. In the commutative case one would then expect a divergent behaviour with power minus two for small $mL$ and an exponential decay for large $mL$. In the present case, even if the small $mL$ behaviour remains the same one, the large limit gets modified to a power-law decay, with power minus three. This can be shown analytically for any $\tilde L$ and seen from the plot in the left panel of Figure \ref{fig.forced1} for the values $\tilde L=0.5$ (red continuous line) and $\tilde L = 3$ (green dashed line). As a consequence of this fact we can say that, as it happens when one introduces interactions \cite{Flachi:2020pvn}, the Casimir force for a massive field in a noncommutative setup is not necessarily exponentially suppressed.}

\coqui{Other regimes arise varying $\tilde L$ for fixed $\tilde m$.  One of them is the large $\tilde L$ limit for fixed $\tilde m$, where the pressure tends to zero, as expected from the definition of Casimir pressure. Another case to discuss  is the (presumably unphysical) one where the distance between the plates is smaller than  $\beta$. As a way to analyze this situation, consider the pressure in the corresponding units of $\beta$, having fixed $\tilde m$. The fact that it tends to a constant for small $\tilde L$ is natural, since $\beta$ acts then as a mass cutoff. One can readily see that this constant equals the second term in the RHS of eq. \eqref{eq:force^C_D=1}, since the first one vanishes.
The same result is also obtained in the strict $L\rightarrow 0$ limit, intended as the first term in the double expansion $\tilde L\ll 1,\,Lm\ll 1$, keeping $\tilde L^{-1}Lm= m\beta$ fixed.
}  We have included in the right panel of Figure \ref{fig.forced1} a plot of the pressure (in the corresponding units of $\beta$) as a function of $\tilde L$, for $\tilde m=1$  (red continous line) and   $\tilde m=0$ (green dashed line).

\coqui{Now we pass to the massless case, where the calculation can be made explicitly}. The \coqui{vacuum pressure then} reduces to
\begin{align}\label{eq:force_D=1_m=0}
 \pressure_{D=1}= \sum_{n=1}^{\infty} \frac{\pi}{8 L^2}\frac{n}{\cosh^2\argo},
\end{align}
and the sum  can be easily evaluated by means of  the Euler-McLaurin formula,
\begin{align}\label{Euler}
\sum_{n=0}^\infty f(n)=\int_0^\infty dn\, f(n)+\left[{1\over2}f(n)+{1\over12}{df(n)\over dn}-{1\over720}{d^3f(n)\over dn^3}+\dots\right]^\infty_0.
\end{align}
The Casimir pressure (like the Casimir energy) is defined by subtracting from this value the contribution in the absence of the plates, which corresponds to the integral in \eqref{Euler}.

Proceeding with the computation, the Euler-McLaurin formula gives rise to a natural expansion in $\beta^2/L^2$, which is the only
dimensionless parameter in the theory and is expected to be very small, since $\beta$ is usually taken of the order of the Planck length.
The explicit computation for \coqui{the Casimir} pressure gives
\begin{align}
\coqui{\pressure^{(C)}_{D=1}}=-\frac{\pi}{96L^2}-\frac{\beta^2\pi^3}{3840L^4}+\mathcal{O}\left(\frac{\beta}{L}\right)^4,
\end{align}
whose first contribution is the usual Casimir pressure, while the second term gives the lowest order correction induced by the Snyder dynamics.

\subsection{On the realization independence in the $D=1$ case}\label{sec:realization}
The results so obtained should be independent from the realization chosen for the anti-Snyder algebra.
We will show this using a different realization of the one-dimensional Snyder algebra. In higher dimensions, an explicit calculation using a different basis can become very involved, because it is difficult to impose the boundary conditions on the plates if one uses realizations of the algebra different from \eqref{eq:realization1}.

Let us consider the Snyder representation \eqref{eq:realization2}.  We will use the symbol $\repre$ to represent the quantities in this particular realization. In one dimension, the operators reduce to
\begin{align}
 \hat{p}_{\repre}=p,\quad \hat{x}_{\repre}=i(1-\beta^2p^2) \partial_p,
\end{align}
acting on functions in a Hilbert space with measure $d\mu=\frac{dp}{1-\beta^2p^2}$ and with $p<\beta^{-1}$.
The momentum eigenstates are trivial,
\begin{align}
 \phi^{\repre}_q(p)=\sqrt{1-\beta^2q^2}\delta(p-q),
\end{align}
while the generalized position eigenstates are
\begin{align}
 \psi^{\repre}_{x_0}(p)=\frac{1}{2\pi}e^{-i\frac{x_0}{\beta} \text{arcth}(\beta p)}
\end{align}
Since they are generalized eigenstates, they can be  normalized according to the formula
\begin{align}
 \int_{p^2<\beta^2} \frac{dp}{1-\beta^2p^2} \psi_{x_0}^{\repre}(p) \psi_{y_0}^{\repre}(p) =\delta(y_0-x_0).
\end{align}

One can now follow the same steps as in Section \ref{sec:spectrum} in order to obtain the spectrum
for the scalar field. One finds that the eigenvalues are quantized and given by
\begin{align}
 \beta q^{\repre}= \tanh\left(\beta\momenta\right),\quad n\in\mathbb{N}^+.
\end{align}
Taking these as the oscillation modes of a scalar field, the Casimir energy is
\begin{align}
 \energy^{\repre}_{D=1}=&\frac{1}{2}\sum_{n=1}^{\infty} \sqrt{{\tanh^2\left(\beta\momenta\right)\over\beta^2}+m^2},
\end{align}
which coincides with eq.~\eqref{eq:energy_D=1}.


\section{The Casimir force in $D=3$}\label{sec:casimir_d=3}
Let us now turn our attention to the physically more relevant case in which the spacetime is given by $\mathcal{M}_{3+1}$.
{Since the integral \eqref{eq:energy_coordinates1} diverges for $q\to\infty$,
we  will regularize it by adding a cutoff $\Lambda_q$ for large momenta coordinates.
This is also natural from a physical point of view, inasmuch as one expects the plates to become transparent in the ultraviolet, generating a natural cutoff. However, a correct UV-cutoff $\Lambda$ of the theory should be defined according to the eigenvalues of the momentum operators, i.e.
\begin{align}\label{eq:cutoff}
\Lambda:=\frac{\Lambda_q}{\sqrt{1+\beta^2\Lambda_q^2}} .
\end{align}
It is clear that $\Lambda<\beta^{-1}$, which is not a sharp constraint since $\beta$ is assumed to be of
the order of the Planck length. In this section we will consider this UV-motivated vision of $\Lambda_q$.
In Section \ref{sec:noncommutative_casimir} we will instead discuss the problem from another perspective, namely the interpretation of $\Lambda_q$ as a geometric quantity, with a role equivalent to that of an IR-cutoff in configuration space.}

\subsection{Massless case}
We start by evaluating the Casimir pressure for a massless field. This problem can be treated in the same way as in one dimension.
Let us consider the expression \eqref{eq:casimir}, which can be cast in the form
\begin{align}\label{Casimener}
 \energy=-\frac{1}{4\pi\beta}\sum_{n=0}^{\infty}\int_0^{\Lambda_q}  dq\, q\,\Delta(q),
 \end{align}
 where
 \begin{align}
 \Delta(q)=\sqrt{1-\frac{1}{(1+\beta^2 q^2)\cosh^2\left(\beta\momenta\right)}}.
 \end{align}

 The \coqui{vacuum pressure} can then be written as
\begin{align}\label{masslesspressure}
 \pressure&=\frac{1}{16L^2}\sum_{n=0}^{\infty}\frac{n\sinh\left(\beta\momenta\right)}{\cosh^3\left(\beta\momenta\right)}
 \int_0^{\Lambda_q}\frac{q\,dq}{(1+\beta^2q^2)\Delta}\nonumber\\
& =\frac{1}{16L^2}\sum_{n=0}^{\infty}\frac{n\sinh\left(\beta\momenta\right)}{\cosh^3\left(\beta\momenta\right)}\left[\ln(2\beta\Lambda_q)
 -\ln\left(1+\tanh\left(\beta\momenta\right)\right)\right]+\mathcal{O}(\Lambda_q^{-1}).
\end{align}
Using  the Euler-McLaurin expansion like in one dimension, and after subtracting the contribution in the absence of the plates, the cutoff $\Lambda_q$ disappears, and one is left with
\coqui{Casimir pressure\footnote{To avoid the proliferation of indices, from now on we will refer to both the vacuum and Casimir pressure with the same symbol, $\pressure$.}}
 \begin{align}\label{masslesforce}
 {\pressure}=-\frac{\pi^2}{7680}\frac{1}{L^4}-\frac{\pi^4 }{48384} \frac{\beta^2}{L^6}+\mathcal{O}(\beta^4).
\end{align}
Again, the first term reproduces the usual Casimir pressure, while the second gives the leading corrections due to the Snyder geometry. Note that the second term
has the same sign as the commutative contribution. Remarkably, in contrast with the commutative case, a finite pressure is obtained by simply subtracting from \eqref{masslesspressure} the vacuum energy in the absence of the plates, without need of further regularization.
\medskip

\subsection{Massive case}
When the field is massive, it is more convenient to use the coordinates introduced in eq.~\eqref{eq:energy_coordinates2}. In these coordinates, $p^2<1/\beta^2$. {We are not going to explicitly write the cutoff $\Lambda$ to simplify the discussion.}

As before, we shall consider the regularized Casimir energy density where the vacuum energy has been subtracted,
\begin{align}
 \energy&=\frac{ \Omega_{D-2}}{2(2\pi)^{D-1}} \int_0^{1/\beta} \frac{dp\, p^{D-2}}{(1-\beta^2p^2)^{D/2+1/2}}\, \left(\sum_{n=1}^{\infty}  \omega_n(p)-\int_{0}^{\infty} dn  \omega_n(p) \right)
\end{align}
with frequencies given by
\begin{align}
\omega_n(p)=\sqrt{p^2+\beta^{-2}\tanh^2\left(\beta\momenta\right) (1-\beta^2p^2) +m^2}.
\end{align}
We have kept track of the dimension $D$ in these equations, in order to render the divergences more visible.

The formal expression for the Casimir pressure is also readily obtained,
\begin{align}\label{eq:force_D=3}
 \pressure&= \frac{\Omega_{D-2}}{2(2\pi)^{D-1}} \int_0^{1/\beta} \frac{dp\, p^{D-2}}{(1-\beta^2p^2)^{D/2-1/2}}\, \left(\sum_{n=1}^{\infty}f_n(p)- \int_0^{\infty} dn\, f_n(p) \right),
\end{align}
where we have introduced the functions
\begin{align}
f_n(p)=  \frac{\momenta}{\beta L}\frac{\tanh\left(\beta\momenta\right)}{\cosh^2\argo\sqrt{p^2+\beta^{-2} \tanh^2\left(\beta\momenta\right)(1-\beta^2p^2)+m^2}}.
\end{align}
The question is once more whether this quantity is regular in $D=3$. Of course, the situation is  more involved that in the one-dimensional case: although there is apparently  only one divergence placed at $p=\beta^{-1}$ (for $D\geq 3$, the expression between parentheses in the RHS of \eqref{eq:force_D=3} could regularize it. Performing an Euler-MacLaurin's expansion, the sum in $n$ is equal to the integral in $n$, up to exponentially vanishing contributions in $\frac{L}{\beta}$,  indicating that the expression \eqref{eq:force_D=3} is not regular.

In any case, as we have seen in the previous sections, it is natural to perform an expansion for small $\beta $, obtaining
\begin{align}\label{eq:energy_D=3_smallb}
 \energy_{D=3}&= \frac{ \Omega_{2}}{8\pi^2}\int_0^{\infty} dp \,p^{} \left(\sum_{n=1}^{\infty}  e_n(p)-\int_{0}^{\infty} dn\;  e_n(p) \right),
\end{align}
where we have defined the quantity
\begin{align}
 e_n(p)=\omega_{n,\beta=0}^{-s} \left[1+ 2 \beta^2 p^2 -\frac{\beta^2}{\omega_{n,\beta=0}^{2}}\left( \frac{p^2\,k_n^2}{2}+\frac{\momenta^4}{4 }\right)  \right].
\end{align}
Note that in these last formulas we have changed the upper limit of the integration in $p$ to $\infty$. This is permitted since in our small $\beta$ expansion we no longer have a divergence in $p=1/\beta$. Moreover, we have employed a $\zeta$-regularization, introducing the $s$ parameter which will be set to $-1$ at the end of the computation.
At this point the procedure follows the commutative one. The integral in the momentum can be explicitly performed and gives
\begin{align}
  \begin{split}
  \int_0^{\infty} dp \,p^{}\, e_n(p)&= \lambda^{-s}\left[\frac{\lambda^{2} }{(s-2) }-\frac{ \beta^2 \momenta^4 }{4 s}-\frac{ \beta^2 \momenta^2}{  (s-2) s } \lambda^{2} +\frac{4 \beta^2 }{s^2-6 s+8 } \lambda^{4}\right],
  \end{split}\\
  \lambda_n&:= \sqrt{\momenta^2+m^2}.
\end{align}
Furthermore, in order to simplify the computations, we trade the sum for integral using the Abel-Plana formula
\begin{align}\label{eq:abel_plana}
 \sum_{n=0}^{\infty} f(n)=\int_0^{\infty} dn f(n) + \frac{1}{2} f(0) + i \int^{\infty}_0 dt \frac{f(i t)-f(-it)}{e^{2\pi t}-1}.
\end{align}
After this step, the first term of the Abel-Plana formula cancels with the integral in the expression \eqref{eq:energy_D=3_smallb} for the Casimir energy density, in which we have regularized subtracting the vacuum contribution. The second term is independent of $L$ and hence irrelevant, since it does not contribute to the pressure. The third is the only relevant one.
After carefully considering the involved functions in the complex plane and setting $s=-1$, we obtain the finite expression
\begin{align}\label{eq:energy_D=3_final}
\begin{split}\energy^{(2)}_{D=3}
&=-\frac{m^3}{12 \pi }+\frac{\beta^2 m^5}{15 \pi }-  \frac{ L m^4}{15\pi^2 } \\
&\hspace{1.5cm}\times \int_{1}^{\infty}  dt\,\frac{\sqrt{t^2-1}}{e^{4 Lm  t}-1}  \left[5 \left(t^2-1\right)+\beta^2 m^2 \left(4 t^4-3 t^2+4\right)\right] .
\end{split}
\end{align}
It is interesting to notice that both noncommutative contributions have the same sign as the commutative one.
This will imply that, at least to this order in $\beta^2$, the overall sign of the Casimir force will be attractive.
In fact, from expression \eqref{eq:energy_D=3_final} one can readily compute the corresponding pressure by taking the derivative with respect to the distance between the plates:
\begin{align}\label{eq:casimir_3d_integral}
 \begin{split}
\pressure^{(2)}_{D=3}&=-\frac{ m^4 }{30 \coqui{\pi^2}  } \int_1^{\infty} dt \,
\frac{\sqrt{t^2-1}}{\left(e^{4 L m t}-1\right)^2}  \left[e^{4 L m t} (4 L m t-1)+1\right]\\
&\hspace{4cm}\times\left[5 \left(t^2-1\right)+\beta^2 m^2 \left(4 t^4-3 t^2+4\right)\right].
 \end{split}
\end{align}

Inasmuch as a closed expression for the integral is not available to us, we proceed to study the large mass and the massless limit.
Unlike the commutative situation where just one dimensionless parameter $m L$ is available,  the regimes of the expression \eqref{eq:energy_D=3_final} are governed also by two other dimensionless parameters, viz.~$\beta m$ and $\beta/L$. However, both of them are small, since $\beta$ is assumed to be of the order of the Planck scale. Curiously, only one of them contributes in the large mass limit of the Casimir pressure \eqref{eq:casimir_3d_integral},
\begin{align}
 \pressure_{D=3}^{(2)}\sim
 -\frac{1}{8 (2\pi)^{3/2}}  \frac{m^{5/2}}{ L^{3/2}} e^{-4 Lm}
\left[1 +\frac{\beta^2 m^2 }{24 }   (32  mL+31)+\mathcal{O}\left((mL)^{-1}\right) \right].
\end{align}
An analog effect is observed also in the massless limit, for which one recovers eq. \eqref{masslesforce}.
Notice that both these results reproduce the commutative case in the limit of vanishing $\beta$. In addition, they show the first noncommutative corrections, which are quadratic in the noncommutativity parameter and of the same sign of the commutative one, thus strengthening the effective pressure.

\begin{figure}
\begin{center}
 \hspace{-1cm}\begin{minipage}{0.49\textwidth}
 \includegraphics[width=1.1\textwidth]{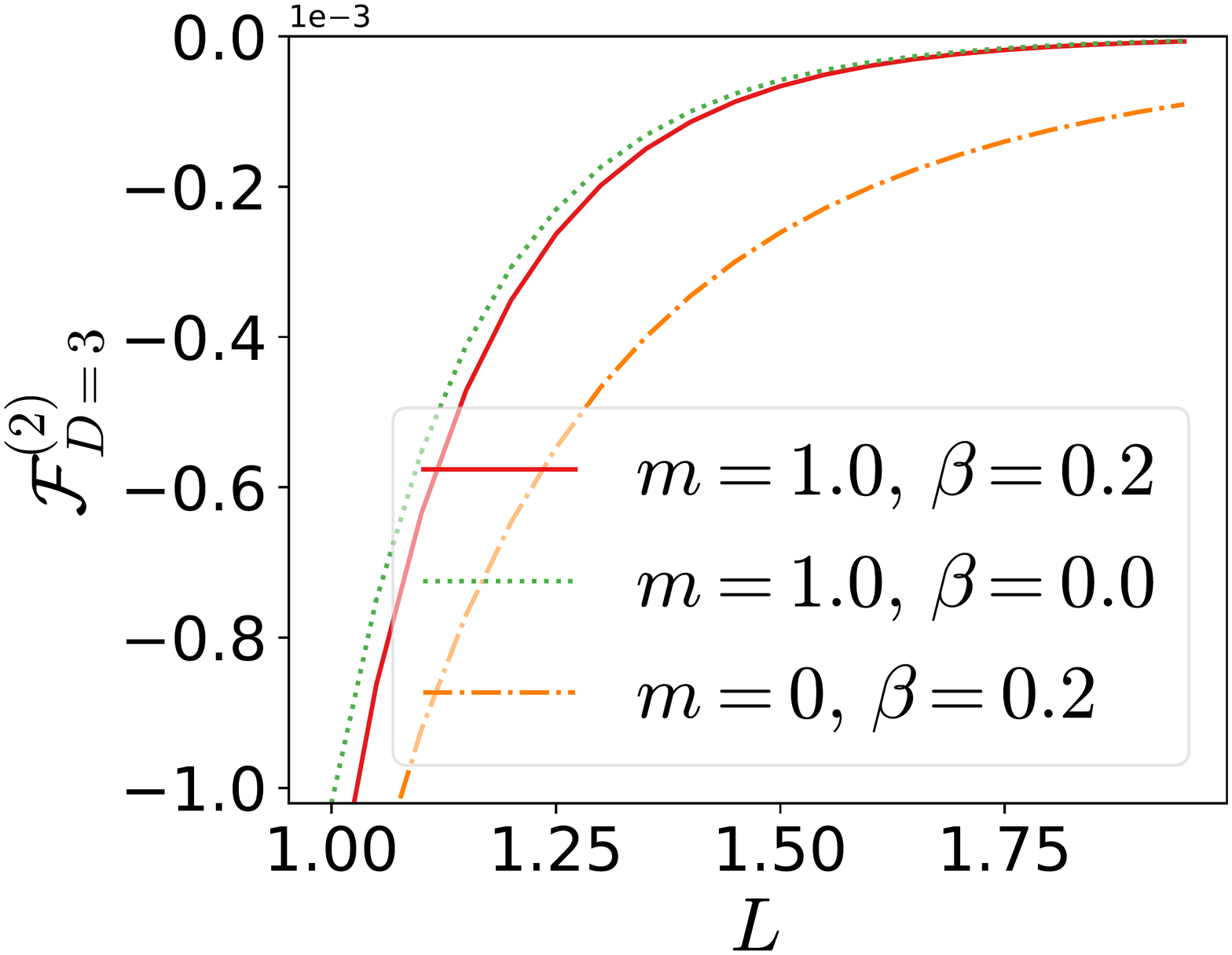}
 \end{minipage}
 \hspace{0.5cm}\begin{minipage}{0.49\textwidth}
 \includegraphics[width=1.1\textwidth]{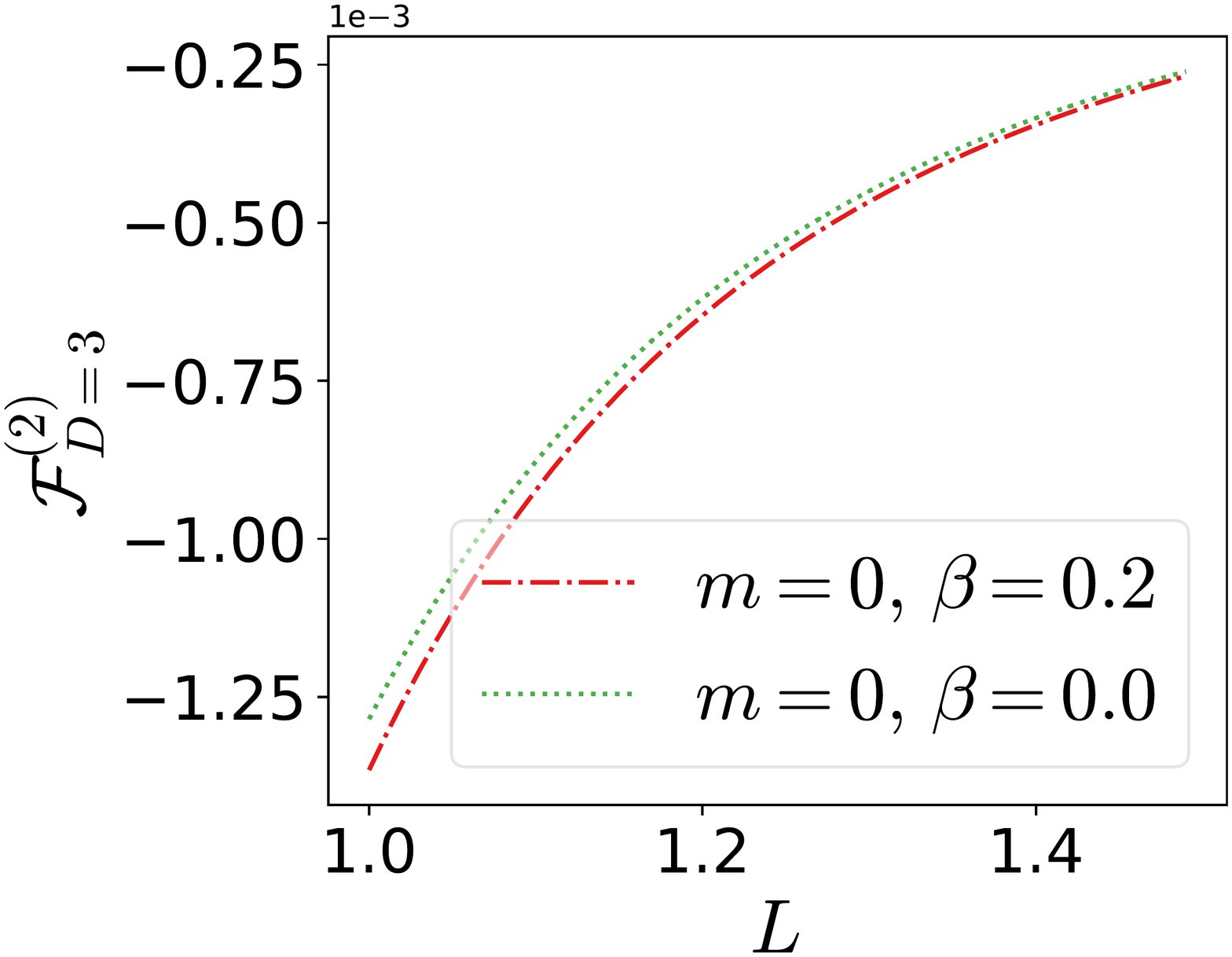}
 \end{minipage}
 \caption{\coqui{Casimir pressure $\pressure^{(2)}_{D=3}$ as a function of the length $L$ for several cases and in arbitrary units. In the left panel, we consider $(m=1,\,\beta=0.2)$ (red continous line), $(m=1,\,\beta=0)$ (green dotted line) and $(m=0,\,\beta=0.2)$ (orange dashed and dotted line), while in the right panel we consider a massless field for $\beta=0.2$ (red dashed and dotted line) and $\beta=0$ (green dotted line).}  }
 \label{fig.forced3}
 \end{center}
\end{figure}
\coqui{One can also perform a numerical integration of expression \eqref{eq:casimir_3d_integral}. In Figure \ref{fig.forced3}, we show the behaviour of the Casimir pressure $\pressure^{(2)}_{D=3}$ as a function of the distance $L$ for several mass and noncommutative parameters in arbitrary units. In the left panel, the exponential decay of the pressure for the massive field can be observed both for the cases of $\beta=0$ (green dotted line) and of $\beta=0.2$ (red continuos line). Moreover, it can be seen that the absolute value of the pressure is greater in the noncommutative case. Also the pressure for a massless field is shown in the left panel (orange dashed and dotted line), to provide a comparison of its power law decay with the previous exponential one.
}

 In the right panel, we plot the pressure for a massless field in a commutative (green and dotted line) and a noncommutative (red dashed and dotted line) setup. Also for a massless field the pressure is larger in the noncommutative case.

\section{{On the goemetric interpretation of $\Lambda_q$ }}\label{sec:noncommutative_casimir}
{Suppose now that there exists no natural UV-cutoff for the plates. In this noncommutative regime where the maximum energy $\beta^{-1}$ could in principle be attained, $\Lambda_q$ in expression \eqref{Casimener} can be thought as a cutoff for long distances in momentum space and, therefore, it can be given
a geometrical meaning. To elaborate on this, let us recall some aspects of the  geometry of the momentum space under consideration.}

Before the inclusion of a confining potential, the spatial momentum space is nothing but the $D$-dimensional hyperbolic space $\mathbb{H}_D$ (or Euclidean AdS$_D$) with radius $\beta^{-1}$, as can be seen from the $\hat{x}$ commutators.  In the particular realization \eqref{eq:realization1}, the volume of $\mathbb{H}_D$ is written as
\begin{align}
\text{Vol(}\mathbb{H}_D)= \Omega_{D-1 } \int_0^{\infty} \frac{dq}{\sqrt{1+\beta^2q^2}}\, q^{D-1}.
\end{align}
Moreover, we can also choose a new coordinate
\begin{align}\label{eq:w}
 \beta w=\text{arcsh}\left(\frac{\beta q_{\perp}}{\sqrt{1+\beta^2q^2_{\parallel}}}\right),
\end{align}
and consider the volume of the hyperplanes of fixed $w$:
\begin{align}
\text{Vol(}\mathbb{H}_{D-1,w=0}):= \Omega_{D-2}\int_0^{\infty} dq\, q^{D-2}.
\end{align}

In order to make contact with our results for the Casimir energy density, recast expression \eqref{eq:casimir} as
\begin{align}\label{eq:casimir_NC}
 \begin{split}
\energy
 &= \frac{\Omega_{D-2}}{2 \beta} \sum_{n=1}^{\infty~}\int_0^{\infty} \frac{dq}{(2\pi)^{D-1}}\,q^{D-2} \sqrt{1+\beta^2 m^2- \frac{1}{(1+\beta^2q^2) \cosh^2\left(\beta\momenta\right)}}\\
 &=\frac{\Omega_{D-2}}{2 \beta} \sum_{n=1}^{\infty~}\int_0^{\infty} \frac{dq}{(2\pi)^{D-1}}\,q^{D-2} \sqrt{1+\beta^2 m^2} \left[1 - \frac{1}{2 u}- \frac{1}{8u^2}+\cdots\right],
 \end{split}
 \end{align}
where $u=(1+\beta^2 m^2)(1+\beta^2q^2) \cosh^2\left(\beta\momenta\right)$. After the expansion, there exists only a finite number of divergent terms in expression \eqref{eq:casimir_NC} for a fixed dimension $D$.
 Using an adequate regularization one can make use of Abel-Plana formula to approximate the series with an integral plus a constant contribution, that in conjunction with the change of variables \eqref{eq:w} gives
\begin{align}
 \frac{\Omega_{D-2}}{2 \beta} \sum_{n=1}^{\infty~}\int_0^{\infty} \frac{dq}{(2\pi)^{D-1}}\,q^{D-2} = \frac{2L}{(2\pi)^{D} \beta }  \text{Vol(}\mathbb{H}_D)  +  \frac{1}{4(2\pi)^{D-1} \beta }  \text{Vol(}\mathbb{H}_{D,w=0}).
\end{align}
This means that in $D<3$ the regularization of the infinities can be done by means of a finite renormalization of the  geometry of momentum space, i.e.~by the inclusion of a momentum-space ``cosmological constant'' and a momentum boundary term of fixed $w$.

In $D\geq3$ the number of divergent terms increases and we have not found a geometrical interpretation of these additional contributions. Although they are apparently given by propagator insertions in the parallel directions with an effective mass given by $\beta^{-1}$, we are not able to pursue further this interpretation.

\section{Conclusions}\label{sec:conclusions}
We have derived the expression for the Casimir energy density of a slab between two parallel plates in an anti-Snyder noncommutative space, working to all orders in the noncommutative parameter $\beta$. This generalizes the computation in Snyder space without boundaries of ref.~\cite{Mignemi:2017yhd}, and also the one in \cite{Casadio:2007ec, Fosco:2007tn} where some heuristic arguments were used in order to give sense to the boundaries.

The divergences encountered during the calculation are milder than in the commutative case. In particular, in the massless case no regularization is needed except the subtraction of the vacuum energy between the plates. Moreover,  the problem presents many interesting theoretical features.
In fact, since the phase space is noncompact, the model possesses an infinite number of eigenstates, in contrast with other models with compact geometries \cite{Chaichian:2001pw}.
However, the effect of the noncommutativity is to impose upper bounds on the physical momenta, i.e.~the momenta of all modes lie inside the sphere $p^2<\beta^{-2}$.

The net effect of such boundedness is to disentangle two divergences that usually appear in the commutative case, viz.~one given by the existence of  modes with momenta as large as desired, and one related to the non-compactness of the momentum space. As stated before, in our case the momenta of the states are bounded, although the geometry of the momentum space,{ the hyperbolic space $\mathbb{H}_D$, is non-compact}. Therefore, the computation of the Casimir energy density, which involves a sum over all modes in momentum space, develops a divergence which should be ascribed to the infinite volume of its geometry.

We have seen that as a consequence, some methods usually employed  to control divergences fail. For example, the use of a $\zeta$-function regularization is precluded by the fact that changing the power to which the energy of the modes is raised in the sum does not help in the convergence. On the other hand, in one dimension the Casimir pressure already yields a well-defined expression.


In the higher-dimensional case, a substantial difference from the $D=1$ instance arises, since a divergence  is present even in the expression for the pressure. A regularization subtracting the vacuum contribution can work in the massless case, but leads to a divergent expression for a massive field. We have then to appeal to a physical cutoff, which prevents the access to energies of the order of $\beta^{-2}$ and allows a small $\beta$ expansion.

Using this expansion, we obtain the first noncommutative corrections to the Casimir pressure for a slab in anti-Snyder space. Their sign is the same as the commutative contribution, thus fostering the effect, as observed in \cite{Harikumar:2019hzq,Frassino:2011aa} for massless particles. Notice that in the massless case our corrections are proportional to $\frac{\beta^2}{L^6}$ as in \cite{Frassino:2011aa}, whereas in \cite{Harikumar:2019hzq} the authors derive for a $\kappa$-Minkowski model, using a different method which entails the introduction of some arbitrary parameters, a contribution proportional to $L^{-4}$. It would be interesting to see if the same method could be applied in (anti-)Snyder space and would give rise to analogous contributions.

\coqui{Another relevant result is related to the large mass behaviour. In the usual commutative situation, one would expect an exponential decay and would therefore neglect the contributions from massive particles to the Casimir effect. However, it is known that this situation may change once interactions are turned on \cite{Flachi:2020pvn}. In our case, we have shown that at least in the $D=1$ free case the noncommutativity leads to a power law decay, $(mL)^{-3}$, hitherto avoiding the mentioned exponential decay. }

In order to gain a deeper insight on the \coqui{geometry of the} Casimir effect, it would also be interesting to study whether there exist other noncommutative spaces in which the geometry of the momentum space and the physical momenta need different regularizations. As mentioned before, in our case the physical momenta (the eigenvalues of momentum operators) are finite, while the geometry of momentum space is noncompact. This means that in order to regularize geometrical expressions one does not need an UV  regularization of the momenta (since physical momenta are already bounded) but rather some kind of  regularization for the noncompactness of  momentum space.  A promising line of investigation could therefore be to further pursue the geometrical analysis initiated in Section \ref{sec:noncommutative_casimir}. This goes in the direction of the momentum-space geometrization program, which has revealed many fundamental features \cite{Carmona:2019fwf, AmelinoCamelia:2011bm}.

Another interesting question is the role that a finite temperature could play in the model under consideration, since several interesting phenomena occur in such regime. Research in these directions is currently carried out.

\medskip
\noindent\textbf{Acknowledgements}:
The authors thank Prof. A.A. Saharian for his useful comments, J. Relancio for profitable discussions and B. Ivetic for his participation in the first stages of this work.
The authors would like to acknowledge networking support by the COST Action CA18108. \coqui{SAF is grateful to G. Gori and the Institut für Theoretische Physik, Heidelberg, for their kind hospitality.}

\printbibliography

\end{document}